\documentclass[runningheads]{cl2emult}

\usepackage{makeidx}  % allows index generation
\usepackage{graphicx} % standard LaTeX graphics tool
\usepackage{subeqnar} % subnumbers individual equations
\usepackage{multicol} % used for the two-column index
\usepackage{cropmark} % cropmarks for pages without
\usepackage{eso}      % placeholder for figures
\makeindex            % used for the subject index
\begin{document}
\title*{Strong constraints on cosmology from galaxy clusters}
\toctitle{Strong constraints on cosmology from galaxy clusters}
% allows explicit linebreak for the table of content
%
%
\titlerunning{Constraints from Clusters}
% allows abbreviation of title, if the full title is too long
% to fit in the running head
%
\author{J.M. Diego\inst{1}
\and E. Mart\'\i nez-Gonz\'alez\inst{1}
\and J.L. Sanz\inst{1}
\and L. Cay\'on\inst{1}
\and J. Silk\inst{2}}
\authorrunning{J.M. Diego et al.}
% if there are more than two authors,
% please abbreviate author list for running head
%
%
\institute{IFCA, CSIC-Univ. Cantabria, Avda. Los Castros, s/n, 39005, Santander, SPAIN
\and Dept. Physics, NAPL, Keble Road, OX1 3RH Oxford, UK}

\maketitle              % typesets the title of the contribution

\begin{abstract}
In this work we show how galaxy clusters can be used to discriminate 
among different cosmological models. We have used available X-ray \& optical 
cluster data to constrain the cosmological parameters as well as the cluster 
scaling relations, $T-M$ and $L_x-T$. We also show the power of future 
SZE data to constrain even more these parameters.
\end{abstract}

\section{Constraints from optical \& X-ray data}
%%%%%%%%%%%%%%%%%%%%%%%%%%%%%%%%%%%%%%%%%%%%%%%%
In several works, authors have used different cluster data sets 
in an attempt to constrain the cosmology. 
The usuall procedure is, starting from the 
Press-Schechter (PS) mass function, fit the experimental 
mass function or by using a given $T-M$ or $L_x-M$ relation build  
some other cluster functions like the temperature, X-ray luminosity or flux 
functions, and then compare them with the corresponding data sets. 
This can be a dangerous process. First, when considering 
just one data set one ensures that its best fitting model is compatible 
with just that data set. 
Some care should be taken to check that the best fitting model is also 
consistent with other data sets.
A second problem comes when the cluster scaling relations $T-M$ 
 or $L_x-M$ are assumed as fixed relations. However, the scatter 
in these relations are known to be important and they can introduce  
uncertainties in the final result. 
In order to avoid all these difficulties we made a fit to different data sets 
simultaneously without doing any assumption about neither the cosmology nor  
the $T-M$ and  $L_x-M$ relations. For the $T-M$ relation we assume the 
free-parameter relation:\\
\begin{equation}
  T_{gas} = T_0 M_{15}^{\alpha}(1 + z)^{\psi}
\end{equation}
where $T_0, \alpha$ and $\psi$ are our three free parameters. 
$M_{15}$ is the cluster mass in $h^{-1} 10^{15} M_{\odot}$ units.
And similarly for the $L_x-M$ :\\
\begin{equation}
  L_x^{Bol} = L_0 M_{15}^{\beta}(1 + z)^{\phi}
\end{equation}
With these two scalings plus the PS formalism we were able to 
build the mass function, the temperature function and the X-ray luminosity 
and flux functions. As can be seen in Diego et al. (2000a) we found 
that only low density universes ($\Omega \approx 0.3$ 
with or without a cosmological constant) are compatible with recent 
determinations of these functions. In that work we also 
obtain some interesting limits for the parameters in the cluster scaling 
relations. Also important is to mention that by combining different data 
sets in our fit, we have reduced significantly the degeneracy in 
$\sigma_8 - \Omega$. In fact, we found a clear peak in the probability 
distribution at the position $\sigma_8 = 0.8, \Omega = 0.3$ 

\section{Constraints from future SZE data}
%%%%%%%%%%%%%%%%%%%%%%%%%%%%%%%%%%%%%%%%%%

The possibilities of cluster data as a cosmological probe will increase 
when SZE surveys will be available. We are studing the possibilities of 
such data sets in order to know how well can we constrain the cosmological 
parameters and how these constraints depend on the assumptions made in, 
for instance, the $T-M$ relation (see Diego et al. 2000b). 
In Diego et al. (2000a) we showed that with present X-ray and optical cluster 
data it is not possible to break the existing degeneracy between low-density 
universes with or without $\Lambda$. In that work we found that both models 
were equally probable when describing the previous data. 
Will it be possible to break this degeneracy with future SZE data ? To answer 
this question we have compared two hypothetical future SZE surveys. 
The first one is based on the Planck satellite. 
This experiment will explore the whole sky at 9 mm frequencies 
(including those where the SZE is more relevant) and with resolutions 
up to 5 arcmin. 
We have estimated that with this experiment it will be possible to 
detect more than 30000 clusters through the SZE which will allow 
to build the curve $N(>S_{mm})$ (see Diego et al. 2000a).
This curve can be used to fit the cosmological models. We have compared 
the $N(>S_{mm})$ curves corresponding to the two best fitting models 
($\Lambda = 0$ and $\Lambda > 0$) found in Diego et al. (2000a). Both models 
predict very similar $N(>S_{mm})$ curves showing being therefore  
difficult to discriminate between the models.
This is not surprising at all, since this curve is dominated by the cluster 
population at low redshift ($z < 0.7$) 
where the degeneracies among the models are more 
important. This point suggests the need of a different data set as a 
cosmological discriminator. 
Apart from the large number of detections, Planck will not provide, 
however, any estimate of the redshift of the clusters.  
In our second experiment we include the redshift of the clusters in order 
to account for evolutionary effects in the cluster population.  

\section{Evolution of the cluster population}
\begin{figure}
\centering
\includegraphics[width=.7\textwidth]{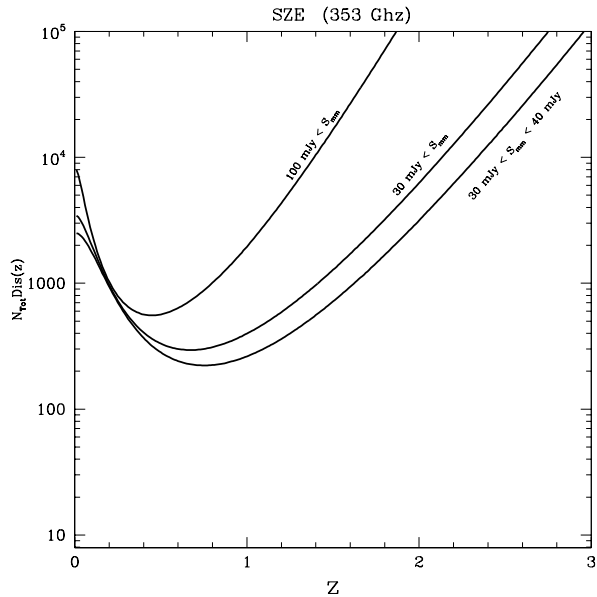}
\caption[]{Number of clusters with measured $z$ required to distinguish 
($3\sigma$) the two models in table 1 in Diego et al (2000a). }
\label{eps1}
\end{figure}

%%%%%%%%%%%%%%%%%%%%%%%%%%%%%%%%%%%%%%%%%%%%%
One advantage of SZE surveys compared with X-ray and optical surveys is 
that the selection function is much less steep (with $z$) in the 
former case. 
Therefore, with the SZE, it seems that we should be able to observe 
deeper in redshift and consequently the information provided by a 
SZE survey would be, {\it a priori}, much more interesting in terms of 
evolution of the cluster population.
Suppose we observe a region of the sky and find 
$N$ SZE detections. Suppose now that we perform optical observations of 
these clusters and obtain their redshifts. Then we have $S_{mm}$ 
and $z$ for each one of the $N$ clusters. 
With this kind of information, how large must be $N$ in order to 
break the degeneracies found in table 1 of Diego et al. (2000a) ?
We have tried to answer these questions by comparing the number of SZE 
detections above a given $z$ for the two degenerated models requiring 
a difference in the models above $3\sigma$ level. 
In Fig.~\ref{eps1}  we show our result.
From this figure we conclude that with only a small subsample of $\sim 300$ 
clusters we could be able to break previous degeneracies. However, as we 
show in Diego et al. (2000b) the best results will come from combining the 
full sky $N(>S_{mm})$ given by Planck and the $N(>z)$ from a small sky patch 
selected from the SZE data. 

%INDEX%%%%%%%%%%%%%%%%%%%%%%%%%%%%%%%%%%%%%%%%%%%%%%%%%%%%%%%%%%%%%%%
\clearpage
\addcontentsline{toc}{section}{Index}
\flushbottom
\printindex
%%%%%%%%%%%%%%%%%%%%%%%%%%%%%%%%%%%%%%%%%%%%%%%%%%%%%%%%%%%%%%%%%%%%%

\end{document}